\documentclass[11pt]{article}\usepackage[margin=1in]{geometry}
 \usepackage[T1]{fontenc}\usepackage{lmodern} \usepackage{microtype}
 \usepackage{mathtools,amssymb,amsthm,bm}\usepackage[dvipsnames]{xcolor}
 \usepackage{graphicx} \usepackage[hidelinks]{hyperref} \begin{document}

 \newcommand\hreff[1]{{\footnotesize\href{https://#1}{https://#1}}}
 \date{} \title {\vspace{-1pc} "Pairs of Squares" Periodic Table} \author
 {Leonid A. Levin\\ \hreff{www.cs.bu.edu/fac/Lnd}\\ Boston University\thanks
 {College of Arts and Sciences, CS, 665 Commonwealth Ave., Boston, MA 02215.}}

 \catcode`\"=12 \maketitle \begin{abstract}\noindent
 I present a new "Pairs of Squares" rendering of the Periodic Table.\\
 It takes advantage of the number of orbitals at each Madelung energy level
being a whole square.\\ This makes the table very uniform and intuitive
in contrast with its currently used presentations. \end{abstract}

In a century and a half since \cite {Mend}, a huge number of forms of
the Periodic Table have been designed (see, e.g., \cite{mzr,int,Scr}).
However, they all share a problem: Their irregularity overwhelms\\
 their periodicity. This is especially bothersome at one's early
exposures to this icon of science.

Yet, there is a numeric pattern none of these layouts seem to exploit
for a full effect. The number of orbitals in each electron shell is a
whole square. And so is their number at each Madelung energy level
(roughly the sum $n+l$ of the first two quantum numbers). And those
squares are each the sum of the first several odd integers, representing
the numbers of orbitals on the respective subshells.

This pattern allows a completely uniform rendering of the table with
a very intuitive look. Each period fills two squares of element cells.
 (Shaped as a single square if entries are $2\times1$ rectangles.)
 Nested subsquares, sharing the "staple" point, group the color-coded
blocks: In each period each subsquare adds one more block in a
$\mathbf\Pi$--shaped stripe. The table fits on a page. But if squares are cut
out and stapled together instead, similar elements fall at the same place
in the respective layers.\footnote
 {As Periods are squares, not lines, element blocks are congruent
 $\mathbf\Pi$-shaped stripes, not segments.\\ Colors make stripes impossible
 to miss. A poor substitute could be adding 1mm space between stripes.}
 A variation has atomic data and elements' symbols instead of full names.
 Another variation has {\bf C}--shaped stripes. On the pdf file,
 entries are clickable, linked to articles about each element.

Of course, one may question the need to add another form of
the Periodic Table to the huge number of those already designed.
But I think one look at this Table (see the next page, in color)
may convince that the extra comfort given by its perfect\footnote
 {Admittedly, a really perfect pattern would start with two
  pink 2-cell periods.\\ However the H/He period is special,
  suggesting a yellow color.\\ So, the small imperfection of merging
  the two into one 4-cell yellow/pink period seems more revealing.}
 regularity comes as some justification.

 \newcommand\NA{-} % \Hd\Cb\Ox\Ph\Sf\RE\Ur
 \newcommand\Hd{{\cell{yellow}{1}{H}{1.01}{Hydrogen}{2.2}{53}{-259}}}
 \newcommand\He{{\cell{yellow}{2}{He}{4}{Helium}{\NA}{31}{-272}}}
 \newcommand\Li{{\cell{myred}{3}{Li}{6.94}{Lithium}{0.98}{167}{180.5}}}
 \newcommand\Be{{\cell{myred}{4}{Be}{9.012}{Beryllium}{1.57}{112}{1287}}}
 \newcommand\B{{\cell{yellow}{5}{B}{10.81}{Boron}{2.04}{87}{2077}}}
 \newcommand\Cb{{\cell{yellow}{6}{C}{12.01}{Carbon}{2.55}{67}{3550}}}
 \newcommand\N{{\cell{yellow}{7}{N}{14.01}{Nitrogen}{3.04}{56}{-210}}}
 \newcommand\Ox{{\cell{yellow}{8}{O}{16}{Oxygen}{3.44}{48}{-219}}}
 \newcommand\F{{\cell{yellow}{9}{F}{19}{Fluorine}{3.98}{42}{-220}}}
 \newcommand\Ne{{\cell{yellow}{10}{Ne}{20.18}{Neon}{\NA}{38}{-249}}}
 \newcommand\Na{{\cell{myred}{11}{Na}{22.99}{Sodium}{0.93}{190}{97.8}}}
 \newcommand\Mg{{\cell{myred}{12}{Mg}{24.31}{Magnesium}{1.31}{145}{650}}}
 \newcommand\Al{{\cell{yellow}{13}{Al}{26.98}{Aluminium}{1.61}{118}{660.3}}}
 \newcommand\Si{{\cell{yellow}{14}{Si}{28.09}{Silicon}{1.9}{111}{1414}}}
 \newcommand\Ph{{\cell{yellow}{15}{P}{30.97}{Phosphorus}{2.19}{107}{44.2}}}
 \newcommand\Sf{{\cell{yellow}{16}{S}{32.06}{Sulfur}{2.58}{105}{115.2}}}
 \newcommand\Cl{{\cell{yellow}{17}{Cl}{35.45}{Chlorine}{3.16}{102}{-102}}}
 \newcommand\Ar{{\cell{yellow}{18}{Ar}{39.95}{Argon}{\NA}{97}{-189}}}
 \newcommand\K{{\cell{myred}{19}{K}{39.1}{Potassium}{0.82}{243}{63.5}}}
 \newcommand\Ca{{\cell{myred}{20}{Ca}{40.08}{Calcium}{1}{194}{842}}}
 \newcommand\Sc{{\cell{myblue}{21}{Sc}{44.96}{Scandium}{1.36}{184}{1541}}}
 \newcommand\Ti{{\cell{myblue}{22}{Ti}{47.87}{Titanium}{1.54}{176}{1668}}}
 \newcommand\V{{\cell{myblue}{23}{V}{50.94}{Vanadium}{1.63}{171}{1910}}}
 \newcommand\Cr{{\cell{myblue}{24}{Cr}{52}{Chromium}{1.66}{166}{1907}}}
 \newcommand\Mn{{\cell{myblue}{25}{Mn}{54.94}{Manganese}{1.55}{161}{1246}}}
 \newcommand\Fe{{\cell{myblue}{26}{Fe}{55.85}{Iron}{1.83}{156}{1538}}}
 \newcommand\Co{{\cell{myblue}{27}{Co}{58.93}{Cobalt}{1.88}{152}{1495}}}
 \newcommand\Ni{{\cell{myblue}{28}{Ni}{58.69}{Nickel}{1.91}{149}{1455}}}
 \newcommand\Cu{{\cell{myblue}{29}{Cu}{63.54}{Copper}{1.9}{145}{1085}}}
 \newcommand\Zn{{\cell{myblue}{30}{Zn}{65.38}{Zinc}{1.65}{142}{419.5}}}
 \newcommand\Ga{{\cell{yellow}{31}{Ga}{69.72}{Gallium}{1.81}{136}{29.8}}}
 \newcommand\Ge{{\cell{yellow}{32}{Ge}{72.63}{Germanium}{2.01}{125}{938.3}}}
 \newcommand\As{{\cell{yellow}{33}{As}{74.92}{Arsenic}{2.18}{114}{817}}}
 \newcommand\Se{{\cell{yellow}{34}{Se}{78.97}{Selenium}{2.55}{103}{221}}}
 \newcommand\Br{{\cell{yellow}{35}{Br}{79.9}{Bromine}{2.96}{94}{-7.3}}}
 \newcommand\Kr{{\cell{yellow}{36}{Kr}{83.8}{Krypton}{3}{88}{-157}}}
 \newcommand\Rb{{\cell{myred}{37}{Rb}{85.47}{Rubidium}{0.82}{265}{39.3}}}
 \newcommand\Sr{{\cell{myred}{38}{Sr}{87.62}{Strontium}{0.95}{219}{777}}}
 \newcommand\Y{{\cell{myblue}{39}{Y}{88.91}{Yttrium}{1.22}{178}{1526}}}
 \newcommand\Zr{{\cell{myblue}{40}{Zr}{91.22}{Zirconium}{1.33}{160}{1855}}}
 \newcommand\Nb{{\cell{myblue}{41}{Nb}{92.91}{Niobium}{1.6}{146}{2477}}}
 \newcommand\Mo{{\cell{myblue}{42}{Mo}{95.95}{{\fns Molybdenum}}{2.16}{139}{2623}}}
 \newcommand\Tc{{\cell{myblue}{43}{Tc}{96.91}{Technetium}{1.9}{136}{2157}}}
 \newcommand\Ru{{\cell{myblue}{44}{Ru}{101.1}{Ruthenium}{2.2}{134}{2334}}}
 \newcommand\Rh{{\cell{myblue}{45}{Rh}{102.9}{Rhodium}{2.28}{134}{1964}}}
 \newcommand\Pd{{\cell{myblue}{46}{Pd}{106.4}{Palladium}{2.2}{137}{1555}}}
 \newcommand\Ag{{\cell{myblue}{47}{Ag}{107.9}{Silver}{1.93}{144}{961.8}}}
 \newcommand\Cd{{\cell{myblue}{48}{Cd}{112.4}{Cadmium}{1.69}{151}{321.1}}}
 \newcommand\In{{\cell{yellow}{49}{In}{114.8}{Indium}{1.78}{167}{156.6}}}
 \newcommand\Sn{{\cell{yellow}{50}{Sn}{118.7}{Tin}{1.96}{158}{231.9}}}
 \newcommand\Sb{{\cell{yellow}{51}{Sb}{121.8}{Antimony}{2.05}{141}{630.6}}}
 \newcommand\Te{{\cell{yellow}{52}{Te}{127.6}{Tellurium}{2.1}{137}{449.5}}}
 \newcommand\I{{\cell{yellow}{53}{I}{126.9}{Iodine}{2.66}{133}{113.7}}}
 \newcommand\Xe{{\cell{yellow}{54}{Xe}{131.3}{Xenon}{2.6}{108}{-112}}}
 \newcommand\Cs{{\cell{myred}{55}{Cs}{132.9}{Caesium}{0.79}{298}{28.4}}}
 \newcommand\Ba{{\cell{myred}{56}{Ba}{137.3}{Barium}{0.89}{253}{727}}}
 \newcommand\La{{\cell{lime}{57}{La}{138.9}{Lanthanum}{1.1}{195}{920}}}
 \newcommand\Ce{{\cell{lime}{58}{Ce}{140.1}{Cerium}{1.12}{182}{798}}}
 \newcommand\Ps{{\cell{lime}{59}{Pr}{140.9}{{\fns Praseodymium}}{1.13}{182}{931}}}
 \newcommand\Nd{{\cell{lime}{60}{Nd}{144.2}{Neodymium}{1.14}{181}{1024}}}
 \newcommand\Pm{{\cell{lime}{61}{Pm}{144.9}{Promethium}{1.13}{183}{1042}}}
 \newcommand\Sm{{\cell{lime}{62}{Sm}{150.4}{Samarium}{1.17}{181}{1072}}}
 \newcommand\Eu{{\cell{lime}{63}{Eu}{152}{Europium}{1.2}{208}{822}}}
 \newcommand\Gd{{\cell{lime}{64}{Gd}{157.2}{Gadolinium}{1.2}{180}{1312}}}
 \newcommand\Tb{{\cell{lime}{65}{Tb}{158.9}{Terbium}{1.1}{175}{1356}}}
 \newcommand\Dy{{\cell{lime}{66}{Dy}{162.5}{Dysprosium}{1.22}{177}{1412}}}
 \newcommand\Ho{{\cell{lime}{67}{Ho}{164.9}{Holmium}{1.23}{176}{1474}}}
 \newcommand\Er{{\cell{lime}{68}{Er}{167.3}{Erbium}{1.24}{175}{1529}}}
 \newcommand\Tm{{\cell{lime}{69}{Tm}{168.9}{Thulium}{1.25}{174}{1545}}}
 \newcommand\Yb{{\cell{lime}{70}{Yb}{173.1}{Ytterbium}{1.1}{194}{824}}}
 \newcommand\Lu{{\cell{myblue}{71}{Lu}{175}{Lutetium}{1.27}{173}{1663}}}
 \newcommand\Hf{{\cell{myblue}{72}{Hf}{178.5}{Hafnium}{1.3}{159}{2233}}}
 \newcommand\Ta{{\cell{myblue}{73}{Ta}{181}{Tantalum}{1.5}{149}{3017}}}
 \newcommand\W{{\cell{myblue}{74}{W}{183.8}{Tungsten}{2.36}{141}{3422}}}
 \newcommand\RE{{\cell{myblue}{75}{Re}{186.2}{Rhenium}{1.9}{137}{3186}}}
 \newcommand\Os{{\cell{myblue}{76}{Os}{190.2}{Osmium}{2.2}{135}{3033}}}
 \newcommand\Ir{{\cell{myblue}{77}{Ir}{192.2}{Iridium}{2.2}{136}{2446}}}
 \newcommand\Pt{{\cell{myblue}{78}{Pt}{195.1}{Platinum}{2.28}{138}{1768}}}
 \newcommand\Au{{\cell{myblue}{79}{Au}{197}{Gold}{2.54}{144}{1064}}}
 \newcommand\Hg{{\cell{myblue}{80}{Hg}{200.6}{Mercury}{2}{149}{-38.8}}}
 \newcommand\Tl{{\cell{yellow}{81}{Tl}{204.4}{Thallium}{1.62}{171}{304}}}
 \newcommand\Pb{{\cell{yellow}{82}{Pb}{207.2}{Lead}{2.33}{175}{327.5}}}
 \newcommand\Bi{{\cell{yellow}{83}{Bi}{208.9}{Bismuth}{2.02}{170}{271.4}}}
 \newcommand\Po{{\cell{yellow}{84}{Po}{209}{Polonium}{2}{168}{254}}}
 \newcommand\At{{\cell{yellow}{85}{At}{210}{Astatine}{2.2}{\NA}{302}}}
 \newcommand\Rn{{\cell{yellow}{86}{Rn}{222}{Radon}{2.2}{120}{-71}}}
 \newcommand\Fr{{\cell{myred}{87}{Fr}{223}{Francium}{0.7}{\NA}{27}}}
 \newcommand\Ra{{\cell{myred}{88}{Ra}{226}{Radium}{0.9}{221}{700}}}
 \newcommand\Ac{{\cell{lime}{89}{Ac}{227}{Actinium}{1.1}{\NA}{1050}}}
 \newcommand\Th{{\cell{lime}{90}{Th}{232}{Thorium}{1.3}{179}{1750}}}
 \newcommand\Pa{{\cell{lime}{91}{Pa}{231}{{\fns Protactinium}}{1.5}{161}{1568}}}
 \newcommand\Ur{{\cell{lime}{92}{U}{238}{Uranium}{1.38}{156}{1132}}}
 \newcommand\Np{{\cell{lime}{93}{Np}{237}{Neptunium}{1.36}{150}{639}}}
 \newcommand\Pu{{\cell{lime}{94}{Pu}{244}{Plutonium}{1.28}{151}{640}}}
 \newcommand\Am{{\cell{lime}{95}{Am}{243}{Americium}{1.3}{173}{1176}}}
 \newcommand\Cm{{\cell{lime}{96}{Cm}{247}{Curium}{1.3}{174}{1340}}}
 \newcommand\Bk{{\cell{lime}{97}{Bk}{247}{Berkelium}{1.3}{170}{986}}}
 \newcommand\Cf{{\cell{lime}{98}{Cf}{251}{Californium}{1.3}{186}{900}}}
 \newcommand\Es{{\cell{lime}{99}{Es}{252}{Einsteinium}{1.3}{\NA}{860}}}
 \newcommand\Fm{{\cell{lime}{100}{Fm}{257}{Fermium}{1.3}{\NA}{1527}}}
 \newcommand\Md{{\cell{lime}{101}{Md}{258}{{\fns Mendelevium}}{1.3}{\NA}{827}}}
 \newcommand\No{{\cell{lime}{102}{No}{259}{Nobelium}{1.3}{\NA}{827}}}
 \newcommand\Lr{{\cell{myblue}{103}{Lr}{262}{Lawrencium}{1.3}{\NA}{1627}}}
 \newcommand\Rf{{\cell{myblue}{104}{Rf}{267}{{\fns Rutherfordium}}{\NA}{\NA}{2100}}}
 \newcommand\Db{{\cell{myblue}{105}{Db}{270}{Dubnium}{\NA}{\NA}{\NA}}}
 \newcommand\Sg{{\cell{myblue}{106}{Sg}{269}{Seaborgium}{\NA}{\NA}{\NA}}}
 \newcommand\Bh{{\cell{myblue}{107}{Bh}{270}{Bohrium}{\NA}{\NA}{\NA}}}
 \newcommand\Hs{{\cell{myblue}{108}{Hs}{269}{Hassium}{\NA}{\NA}{\NA}}}
 \newcommand\Mt{{\cell{myblue}{109}{Mt}{278}{Meitnerium}{\NA}{\NA}{\NA}}}
 \newcommand\Ds{{\cell{myblue}{110}{Ds}{281}{{\fns Darmstadtium}}{\NA}{\NA}{\NA}}}
 \newcommand\Rg{{\cell{myblue}{111}{Rg}{281}{{\fns Roentgenium}}{\NA}{\NA}{\NA}}}
 \newcommand\Cn{{\cell{myblue}{112}{Cn}{285}{Copernicium}{\NA}{\NA}{\NA}}}
 \newcommand\Nh{{\cell{yellow}{113}{Nh}{286}{Nihonium}{\NA}{\NA}{\NA}}}
 \newcommand\Fl{{\cell{yellow}{114}{Fl}{289}{Flerovium}{\NA}{\NA}{\NA}}}
 \newcommand\Mc{{\cell{yellow}{115}{Mc}{289}{Moscovium}{\NA}{\NA}{\NA}}}
 \newcommand\Lv{{\cell{yellow}{116}{Lv}{293}{Livermorium}{\NA}{\NA}{\NA}}}
 \newcommand\Ts{{\cell{yellow}{117}{Ts}{293}{Tennessine}{\NA}{\NA}{\NA}}}
 \newcommand\Og{{\cell{yellow}{118}{Og}{294}{Oganesson}{\NA}{\NA}{\NA}}}
 \newcommand\Uu{{\cell{myred}{119}{Uu}{\NA}{Ununennium}{\NA}{\NA}{\NA}}}
 \newcommand\Ub{{\cell{myred}{120}{Ub}{\NA}{Unbinilium}{\NA}{\NA}{\NA}}}

 \newcommand\rsc  {https://periodic-table.rsc.org/element/}
 \newcommand\pbcm {https://pubchem.ncbi.nlm.nih.gov/element/}
 \newcommand\wbel{https://www.webelements.com/} \definecolor{pale}{rgb}{1,1,.7}
 \definecolor{myred}{rgb}{1,.6,.6} \definecolor{myblue}{rgb}{.5,.8,1}
 \newcommand\fnt[1]{\setlength\fsz{#1pt}\fontsize{\fsz}{1.2\fsz}\selectfont}
 \newlength\fsz \newcommand\fns{}

 \newcommand\prda{} \newcommand\prdb{} \newcommand\prdc{} \newcommand\prdd{}
 \newcommand\prde{} \newcommand\prdf{} \newcommand\prdg{} \newcommand\prdh{}
 \newcommand{\prd}[1]{\ifcase#1 \prda \or\prdb \or\prdc \or\prdd
   \or\prde \or\prdf \or\prdg \or\prdh\fi} \newcommand\tbl{}
 \newlength{\bw}\newcommand\bx[2][c] {{\settowidth{\bw}
   {\begin{tabular}{@{}l@{}}#2\end{tabular}}\parbox[#1]{\bw}{#2}}}

 \newcommand\fprd{H/He period is yellow, thus\\its placement is slighty odd.}
 \newcommand{\stpl} {{\renewcommand{\arraystretch}{.3}\bf \begin{tabular}
 {@{}c@{}}\rule{8pt}{0pt}\\[-4pt]S\\t\\a\\p\\l\\e\\[-2pt]\end{tabular}}}
 \newcommand{\PSPT}{"Pairs of Squares"} \newcommand{\rndr} {This {\PSPT}
 colored rendering of the Periodic Table seems ~most intuitive in view of
 quadratic number of orbitals at each atomic energy level.}
 \newpage \newgeometry{margin=2pc}\fnt{10}\begin{center}

 \renewcommand\prda{\Hd\\\He} \renewcommand\prdb{\Li\Be}
 \renewcommand\prdc{\Cb\N\Ox\F\\ \B\Na\Mg\Ne}
 \renewcommand\prdd{\Si\Ph\Sf\Cl\\ \Al\K\Ca\Ar}
 \renewcommand\prde{\V\Cr\Mn\Fe\Co\Ni\\ \Ti\Ge\As\Se\Br\Cu\\
  \Sc\Ga\Rb\Sr\Kr\Zn} \renewcommand\prdf{\Nb\Mo\Tc\Ru\Rh\Pd
  \\ \Zr\Sn\Sb\Te\I\Ag\\ \Y\In\Cs\Ba\Xe\Cd}
 \renewcommand\prdg{\Nd\Pm\Sm\Eu\Gd\Tb\Dy\Ho\\ \Ps\Ta\W\RE\Os\Ir\Pt\Er
  \\ \Ce\Hf\Pb\Bi\Po\At\Au\Tm\\ \La\Lu\Tl\Fr\Ra\Rn\Hg\Yb}
 \renewcommand\prdh{\Ur\Np\Pu\Am\Cm\Bk\Cf\Es\\ \Pa\Db\Sg\Bh
   \Hs\Mt\Ds\Fm\\ \Th\Rf\Fl\Mc\Lv\Ts\Rg\Md\\ \Ac\Lr\Nh\Uu\Ub\Og\Cn\No}
 \renewcommand{\tbl}[1] {{\begin{tabular}[b]{c}\prd{#1}\\
       \fbox{\bx{#1. staple here}}\end{tabular}}}

 \newcommand{\cell}[8]{\lowercase{\def\lcname{#5}}\fbox
 {\href {\wbel\lcname} {\colorbox{#1}{\parbox[b] %{\rsc#2} %{\pbcm#2
 [52pt][c]{24pt} {{\fne#2~#3}\\#4\\#6\\#7\\#8}}}}} \newcommand\fne{\fnt9}

 \renewcommand\prdb
 {\bx[b]{\colorbox{yellow}{\parbox[b][58pt]{66pt}\fprd}\\\Li\Be}}
 \raisebox{16pt}{\bx[t]{\colorbox{yellow}{\rule{0pt}{121pt}
 \bx[b]{ ~ ~ }}}}\tbl{1}\hspace{-3pt}\bx[b]
 {\prd0\\\hspace*{10pt}\fbox{0.}} \hspace{4pc} \tbl{2} \tbl{3}\\[6pt]

 \mbox{\bx{\null\hfill\fcolorbox{black}{pale}{\parbox[b]{195pt}
 {\fnt{12}\bf Agenda: number (Z), Symbol, Weight, Electronegativity,
 Atomic Radius (pm), Melting point (C$^o$).}} \hfill\null\\[16pt]
 \tbl{4}\\[1pc]\null\hfill\fcolorbox{black}{pale}
 {\parbox[b]{16pc}{\rndr~(Stripes are $\mathbf\Pi$-shaped.)}}
 \hfill\null\\[8pt] \tbl{5}}}\rule[-22pc]{2pt}{480pt}
 \bx{\tbl{6}\\[8pt] \renewcommand\fne{\fnt7} \tbl{7}}\\[6pt]
 {\large\bf{\PSPT} Periodic Table. (Data, $\mathbf\Pi$-shaped Stripes).}

 \newpage \renewcommand\prda{\Hd\\\He} \renewcommand\prdb{\Li\Be}
 \renewcommand{\cell}[8]{\lowercase{\def\lcname{#5}}\fbox
 {\href {\wbel\lcname} {\colorbox{#1}{\parbox[b]
 [18pt][c]{55pt}{#2 #3 #4\\{\renewcommand\fns{\fnt8}#5}}}}}}

 \bx{\hspace*{2pc}\fcolorbox{black}{pale}{\parbox[b]{16pc}
  {\rndr~(Stripes are $\mathbf\Pi$-shaped.)}}\\[24pt]
 \raisebox{14pt}{\colorbox{yellow}{\rule{0pt}{56pt} ~~}}\hspace{-6pt}
 \bx[b]{ ~ \colorbox{yellow}{\parbox{127pt}\fprd}\\[5pt]\tbl{1}}
 \hspace{-6pt}\bx[b] {\prd0\\\hspace*{27pt}\fbox{0.}}}
  ~ \bx{\rule[18pt]{2pt}{2in}}~\bx{\tbl{2}\\[5pt]\tbl{3}}\\[5pt]
 \tbl{4}\\[6pt] \tbl{5}\\[6pt] {\tbl{6}\\[6pt]\tbl{7}}\\[1pc]
 {\large\bf{\PSPT} Periodic Table ($\mathbf\Pi$-shaped Stripes).}

 \newpage \renewcommand\prda {\Hd\He} \renewcommand\prdb {\Li\\ \Be}
 \renewcommand\prdc {\Cb\B\\ \N\Na\\ \Ox\Mg\\ \F\Ne}
 \renewcommand\prdd {\Si\Al\\ \Ph\K\\ \Sf\Ca\\ \Cl\Ar} \renewcommand\prde
  {\V\Ti\Sc\\ \Cr\Ge\Ga\\ \Mn\As\Rb\\ \Fe\Se\Sr\\ \Co\Br\Kr\\ \Ni\Cu\Zn}
 \renewcommand\prdf
  {\Nb\Zr\Y\\ \Mo\Sn\In\\ \Tc\Sb\Cs\\ \Ru\Te\Ba\\ \Rh\I\Xe\\ \Pd\Ag\Cd}
 \renewcommand\prdg {\Nd\Ps\Ce\La\\ \Pm\Ta\Hf\Lu\\ \Sm\W\Pb\Tl\\
 \Eu\RE\Bi\Fr\\ \Gd\Os\Po\Ra\\ \Tb\Ir\At\Rn\\ \Dy\Pt\Au\Hg\\ \Ho\Er\Tm\Yb}
 \renewcommand\prdh {\Ur\Pa\Th\Ac\\ \Np\Db\Rf\Lr\\ \Pu\Sg\Fl\Nh\\
 \Am\Bh\Mc\Uu\\ \Cm\Hs\Lv\Ub\\ \Bk\Mt\Ts\Og\\ \Cf\Ds\Rg\Cn\\ \Es\Fm\Md\No}
  {\renewcommand{\tbl}[1] {\bx{\prd{#1}}\bx{\fbox{#1.}\\\fbox\stpl}}

 \bx{\colorbox{pale}{\rule[-4pt]{0pt}{18pt}Merged H/He, Li/Be periods:}\\
 \mbox{\colorbox{pale}{\parbox[c]{62pt}{\raggedright~\\[4pt]
 \fprd\\[4pt]}}}\tbl{1}\\[1pt]\prd{0}\fbox{0.}}
 \hspace{3pc} \tbl{2} ~ \tbl{3}\\[1pc]

 \mbox{\bx{\tbl{4}\\[1pc]\fcolorbox{black}{pale}{\parbox[b]{16pc}
 {\rndr~(Stripes are C-shaped.)}}\\[1pc]\tbl{5}}
 \bx{~ \rule{2pt}{38pc}}} ~ \bx{\tbl{6}\\[1pc]\tbl{7}}
 \\[1pc]{\large\bf{\PSPT} Periodic Table (C-shaped Stripes).\\}}

\end{center} \end{document}